\documentclass[5p]{elsarticle}

\usepackage{lineno, hyperref}
\usepackage{dblfloatfix} 
\usepackage{float}
\usepackage{graphicx}
\usepackage{soul} 
\usepackage{xcolor} 
\usepackage{caption}
\usepackage{bm}
\usepackage{svg}
\usepackage{amsmath}
\usepackage[normalem]{ulem}
\usepackage{upgreek}
\usepackage{tikz}
\usepackage{tabularx}

\usepackage{nicefrac}

\journal{ArXiv}

\begin{document}
\nolinenumbers

\begin{frontmatter}

\title{Irreversible evolution of dislocation pile-ups during cyclic microcantilever bending}

\author[ELTEaddress]{D\'{a}vid Ugi}
\author[KITaddress,HKAaddress]{Kolja Zoller}
\author[ELTEaddress]{Kolos Luk\'{a}cs}
\author[MFAaddress]{Zsolt Fogarassy}
\author[ELTEaddress]{Istv\'{a}n Groma}
\author[LGFaddress]{Szilvia Kal\'{a}cska\corref{mycorr}}\cortext[mycorr]{Corresponding authors} \ead{szilvia.kalacska@cnrs.fr}
\author[KITaddress,HKAaddress]{Katrin Schulz\corref{mycorr}} \ead{katrin.schulz@kit.edu}
\author[ELTEaddress,FZJaddress]{P\'{e}ter Dus\'{a}n Isp\'{a}novity\corref{mycorr}}\ead{ispanovity.peter@ttk.elte.hu}

\address[ELTEaddress]{E\"{o}tv\"{o}s Lor\'{a}nd University, Department of Materials Physics, P\'{a}zm\'{a}ny P\'{e}ter s\'{e}tany 1/a, 1117 Budapest, Hungary}
\address[KITaddress]{Karlsruhe Institute of Technology, Institute for Applied Materials (IAM), Kaiserstr. 12, 76131 Karlsruhe, Germany}
\address[HKAaddress]{Hochschule Karlsruhe - University of Applied Sciences (HKA), Moltkestr. 30, 76133 Karlsruhe, Germany} 
\address[MFAaddress]{Centre for Energy Research, Institute of Technical Physics and Materials Science, Konkoly Thege M. \'{u}t 29-33., 1121 Budapest, Hungary}
\address[LGFaddress]{Mines Saint-Etienne, Univ Lyon, CNRS, UMR 5307 LGF, Centre SMS, 158 cours Fauriel 42023 Saint-\'{E}tienne, France}
\address[FZJaddress]{Institute for Advanced Simulation: Materials Data Science and Informatics (IAS-9), Forschungszentrum J\"{u}lich GmbH, 52425 J\"{u}lich, Germany}

\begin{abstract}
In metals geometrically necessary dislocations (GNDs) are generated primarily to accommodate strain gradients and they play a key role in the Bauschinger effect, strain hardening, micron-scale size effects and fatigue. During bending large strain gradients naturally emerge which makes this deformation mode exceptionally suitable to study the evolution of GNDs. Here we present bi-directional bending experiment of a Cu single crystalline microcantilever with in situ characterisation of the dislocation microstructure in terms of high-resolution electron backscatter diffraction (HR-EBSD). The experiments are complemented with dislocation density modelling to provide physical understanding of the collective dislocation phenomena. We find that dislocation pile-ups form around the neutral zone during initial bending, however, these do not dissolve upon reversed loading, rather they contribute to the development of a much more complex GND dominated microstructure. This irreversible process is analysed in detail in terms of the involved Burgers vectors and slip systems to provide an in-depth explanation of the Bauschinger-effect and strain hardening at this scale. We conclude that the most dominant role in this behaviour is played by short-range dislocation interactions.
\end{abstract}


\end{frontmatter}

\section{Introduction}

Since the production of micron- and submicron-scale samples using a focused ion beam (FIB) and their subsequent mechanical testing became a state-of-the-art technology, the field of micromechanics has seen a rapid development. The main reason for the increased interest towards this field is three-fold: Firstly, materials at this scale exhibit remarkable features and often superior mechanical properties compared to their bulk counterparts. For instance, they generally show the `smaller is harder' type of size effects \cite{uchic2004sample, greer2005size, uchic2009plasticity} and may exhibit unpredictable avalanche-like behaviour \cite{dimiduk2006scale, zaiser2006scale, ispanovity2022dislocation}. Secondly, the small volume of the sample provides unique opportunities to observe specific deformation mechanisms using various experimental techniques to unprecedented detail \cite{shan2008mechanical, kiener2011work, samaee2021deciphering, mathis2021dynamics}. Finally, at the micron-scale it becomes feasible to model the evolution of the whole microstructure with various physics based computational methods \cite{csikor2007dislocation, zoller2021microstructure}. For these reasons, a range of different types of micromechanical experiments have been suggested and conducted, such as, microcompression \cite{dimiduk2005size, greer2005size, volkert2006size}, microtension \cite{kiener2008further, kim2009tensile, kirchlechner2011dislocation}, microshear \cite{wieczorek2016assessment, guillonneau.2022}, microtorsion \cite{fu.2016}, microbending \cite{motz2005mechanical, gong2009anisotropy} and microfatigue \cite{Lavenstein.2019, iyer2021cyclic}.

In this paper the focus is on the plasticity of single crystalline metals where irreversible deformation is dominated by the collective dynamics of lattice dislocations. Here, with respect to the specimen size, two regimes of different behaviour can be distinguished. In the so-called \emph{dislocation starved regime}, when the sample dimension is below approx.\ 1~$\upmu$m, the number of  dislocation lines and dislocation sources is so small, that dislocation interactions (such as annihilation, junction formation or cross-slip events) hardly take place and the plastic response and size effects are dominated by individual dislocation mechanisms (such as source activation) \cite{greer2005size, shan2008mechanical}. In the so-called \emph{storage regime}, on the other hand, dislocation interactions do happen in an increasingly large amount during deformation, leading to the build-up of a complex dislocation network and the increase of the dislocation density \cite{zhao2019critical, kiener2011work, kirchlechner2011dislocation, kalacska2020investigation}. The observed size effects and strain hardening can be often related to the development of strain gradients and the corresponding geometrically necessary dislocation (GND) content. Although such structures were found to develop even during uniaxial microcompression tests \cite{zoller2021microstructure}, microcantilever bending is a more natural choice to study the effect of strain gradients and the developments of GNDs since for this deformation mode strain gradients are inherently present already in the elastic regime. It is noted, that for the same reason microcantilever bending has been used successfully to study the role of strain gradients in materials with even more complex microstructure, see, e.g. \cite{Kalacska.2019, Kreins.2021, Wimmer.2015, Kapp.2017, Alfreider.2018}.

The first microcantilever bending tests were carried out by Motz et al.~with a focus on the multiple slip deformation of copper single crystals \cite{motz2005mechanical}. The experiments were later complemented by discrete dislocation dynamics (DDD) simulations \cite{motz2008micro}. Both approaches suggested that plasticity and size effects are dominated by the development of individual dislocation pile-ups close to the neutral zone. These polarized dislocation structures also give rise to a pronounced Bauschinger effect (that is, asymmetry of the flow stress w.r.t.\ the loading direction) \cite{kapp2015importance} and induce a strong internal back stress that, for instance, has a significant influence on crack propagation during fatigue \cite{eisenhut2017effect}. DDD simulations in 2D \cite{tarleton2015discrete} and 3D \cite{motz2008micro, velayarce2018influence} confirmed the formation of the pile-ups arranged symmetrically around the neutral zone.
The role of temperature and crystal orientation on the formation of pile-ups and the corresponding size effects were also investigated experimentally \cite{velayarce2018influence}.

Most microbending experiments mentioned so far focused on unidirectional loading and the formation and role of pile-ups during bending. However, it is equally important to investigate the evolution of the microstructure during reversed loading to understand the role of GNDs and strain hardening during fatigue. Plastic deformation and the corresponding dislocation motion is generally considered irreversible, however, under certain conditions it can, in fact, be reversible. A prominent example is a sequence of dislocations emitted from the same source, therefore lying on the same slip plane and piling up against an obstacle. Upon reversed loading dislocations in such a pile-up may return to their origin and annihilate, thus recovering the exact original atomic arrangement. This process was confirmed by \emph{in situ} micro Laue diffraction experiments performed during low cycle bending fatigue of single crystalline copper microcantilevers with a cross section of approx.\ $10 \times 10 \, \upmu$m$^2$ \cite{kirchlechner2015reversibility}. Generally speaking, broadening of X-ray diffraction peaks in single crystals are caused by internal misorientations due to GNDs. Kirchlechner \emph{et al.}\ found that peak broadening and the formation of sub-peaks during microbending was to a large extent reversible upon unloading \cite{kirchlechner2015reversibility}. This means that GND storage or cell formation was not observed at low cycle numbers in spite of the relatively large sample size. However, several possible origins of irreversibility were discussed such as cross-slip and dislocation–dislocation interactions with the formation of new Burgers vectors (BVs, e.g., Lomer locks) \cite{kirchlechner2015reversibility}. Stricker \emph{et al.}~also discussed a possible mechanism of irreversibility during which the dissolution of the pile-up after unloading leaves behind a slip step on the surface \cite{stricker2017irreversibility}.

In order to understand the formation of pile-ups and to tackle the issue of plastic reversibility/irreversibility upon reversed loading, in this paper bending of Cu microcantilevers is studied. To promote dislocation reactions, a multiple slip orientation is chosen. The evolution of the microstructure is analysed mainly using \emph{in situ} high-resolution electron backscatter diffraction (HR-EBSD), a cross-correlation based method that provides detailed microstructural information based on the Kikuchi patterns \cite{arsenlis1999crystallographic, wilkinson2010determination}. In particular, components of the stress and distortion tensor, as well as three components of the Nye dislocation density tensor can be measured on the surface of the sample with a spatial resolution of about 100 nm. As such, this technique is especially useful for characterizing dislocation structures that develop close to the surface \cite{kalacska2020_3d}. The method can also be applied for three-dimensional (3D) reconstruction of the microstructure using FIB serial sectioning, this, however, leads to the destruction of the sample, see, e.g, \cite{kalacska2020investigation}. The method was also applied for a Cu microcantilever to measure the 3D GND structure after unidirectional bending and a quite homogeneous distribution of GNDs along the cross section of the beam was found \cite{Konijnenberg.2015}.

To gain deeper physical insights into the dislocation mechanisms during microbending and to explain the findings of the HR-EBSD measurements, experiments are complemented with continuum dislocation density (CDD) simulations \cite{hochrainer2007three,hochrainer.2014,hochrainer2015multipole, schulz.2019}. The advantage of this method compared to DDD simulations is that a larger volume can be modelled up to larger strains, yet, this method, contrary to even higher scale crystal plasticity approaches, can account for every individual dislocation interaction that can take place in the FCC crystal. In addition, CDD is able to model all the quantities measured by HR-EBSD, so, comparison of experiments and modelling can be done in an explicit manner, as was shown earlier for microcompression experiments \cite{zoller2021microstructure}. As such, this method is very well suited for the aim of the present paper, that is, to understand to what extent is pile-up formation reversible, and to identify the dislocation mechanisms that are responsible for possible irreversibilities.

The paper is organised as follows. This introduction is followed by a brief crystallographic analysis of the microbending studied to make general predictions on the slip system activities. Then, section \ref{sec:methods} introduces the experimental methods and the CDD framework followed by a detailed overview of the results obtained by both experiments and modelling in section \ref{sec:results}. Section \ref{sec:discussion} provides an in-depth discussion of the findings and concludes the paper with a short outlook.

\section{System analysis}
\label{sec:system_analysis}

\begin{figure}[!h]
    \includegraphics[width=0.48\textwidth]{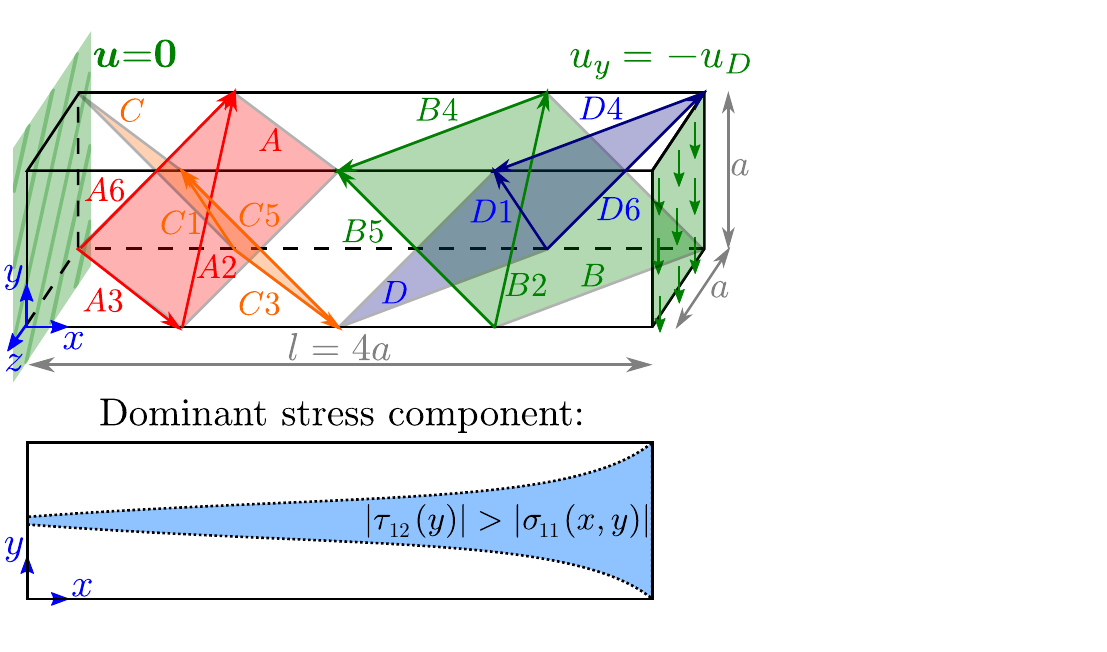}
    \centering
    \begin{picture}(0,0)
    \put(48,65){\begin{minipage}{5cm}\small 
    Slip plane normals:\vspace*{0.15cm} \\
    A: $(1\bar{1}\bar{1})$ \,\,\, B: $(111)$\\
    C: $(11\bar{1})$ \,\,\, D: $(1\bar{1}1)$\\    
    
    \noindent
    Slip directions:\vspace*{0.15cm} \\
    1: $[011]$ \,\,\, 2: $[01\bar{1}]$\\
    3: $[101]$ \,\,\, 4: $[\bar{1}01]$\\
    5: $[\bar{1}10]$ \,\,\, 6: $[110]$
    \end{minipage}
    }
    \put(-120,145){\sffamily{a)}}
    \put(48,121){\sffamily{b)}}
    \put(-120,64){\sffamily{c)}}
    \end{picture}    
    \caption{a) Considered system geometry and the slip planes of the FCC sample. b) Specific vectors corresponding to the slip planes and the BVs. c) The schematics highlights with blue the region on the front face of the cantilever where the shear stresses are larger than the compressive stresses.
    \label{fig:slip_systems}
    }
\end{figure}
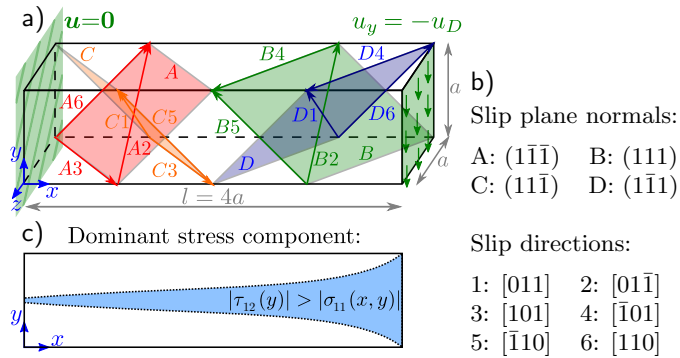

In this paper the plastic bending of a Cu single crystalline microcantilever of rectangular shape and square cross section is investigated. The cantilever is oriented for $\langle 100 \rangle \{ 100 \}$ multiple slip, that is, edges of the beam are parallel with the edges of the cubic unit cell of the FCC lattice. The sketch of the geometry is shown in Fig.~\ref{fig:slip_systems}a where, according to the Schmid-Boas notation \cite{schmid.1935}, the four individual slip planes are denoted by letters A--D and the slip directions (or BV directions) by numerals 1--6 (specified in Fig.~\ref{fig:slip_systems}b). On the left side, the end of the beam is fixed, that is, no displacements are allowed, whereas a displacement $u_\mathrm{D}$ is prescribed at the right end of the cantilever that results in force $F$. Free boundary conditions apply on the rest of the sides. Since the cantilever is initially bent downwards, positive $u_\mathrm{D}$ and $F$ values refer to displacements and forces in the negative $y$ direction. In both experimental and simulation studies cantilever size is chosen as $a=5\,{\upmu}$m that, according to the Introduction above, is expected to be at the limit of the starved and storage regimes.

In this section we aim to make predictions about the plastic activities on the individual slip systems based on the crystal orientation and deformation mode. To this end, we, first, analyse the stresses developing during elastic bending of the cantilever. The bending moment along the $z$ axis $M(x)=F(x-l)$ (with $l$ being the length of the cantilever or, equivalently, the moment arm) is linear along the $x$-axis and, consequently, the normal and the shear stresses read as
\begin{equation}
    \sigma_{11}(x,y)=-\frac{12 M(x)}{a^4}\left( y-\frac{a}{2} \right)
    \label{eqn:sigma_xx}
\end{equation}
and
\begin{equation}
    \tau_{12}(y)=-\frac{6\,F}{a^2}\left[\frac{1}{4}-\left(\frac{y}{a}-\frac{1}{2}\right)^2\right],
\end{equation}
respectively. Figure~\ref{fig:slip_systems}c shows qualitatively the region where the magnitude of the shear stress exceeds that of the normal stress. At the fixed end of the beam, where the largest stresses occur, this region is rather thin because the ratio of the maxima of the occurring normal and shear stresses is quite large: $\frac{|\sigma_{11}|_\mathrm{max}}{|\tau_{12}|_\mathrm{max}}=4 \cdot l/a$. Accordingly, the shear stress is expected to play an inferior role during bending.


Based on this analysis one can group slip systems into four categories: \emph{Group I} contains slip systems where both normal and shear stresses contribute to the resolved shear stress (RSS) (A3, B4, C3, D4), \emph{Group II} contains the ones where only normal stresses have contributions (A6, B5, C5, D6) and \emph{Group III} comprises systems with only shear stresses playing a role (A2, B2, C1, D1). The distinction of slip systems of Group I and Group II is also motivated by the fact, that BVs 3 and 4 lie in the $xz $ plane, so a single dislocation moving to the neutral zone during bending will have a pure screw character. As a result, dislocations in Group I are expected to cross-slip at a larger rate compared to the other two groups that may contribute significantly to irreversibility.

\section{Methods}\label{sec:methods}

\subsection{Experimental}

\begin{figure}[!h]
    \centering
    
    \includegraphics[width=0.4\textwidth]{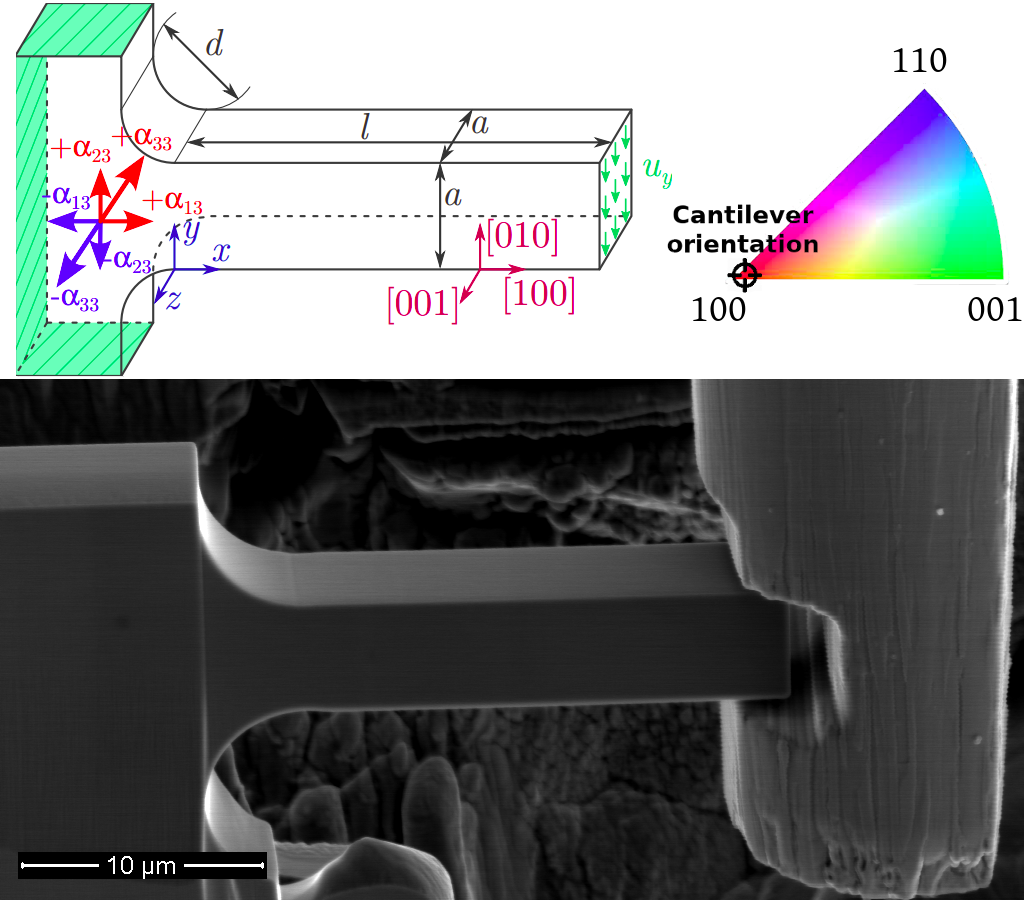}
    \put(-215,170){\sffamily{a)}}
    \put(-70,170){\sffamily{b)}}
    \put(-205,94){\sffamily{\textcolor{white}{c)}}}
    \caption{Experimental conditions to achieve bi-directional bending deformation. (a): The intended geometry and crystal orientation as well as the definition of the coordinate system used throughout the paper. (b): Orientation of the undeformed cantilever indicated on the inverse pole figure (IPF) triangle based on EBSD measurements. The cantilever was found to be of a single crystal with a slight $5.2^\circ$ mismatch between the actual and the preferred orientation along the $z$ axis. (c): SEM image of the cantilever before testing. The custom made W bending tool is also shown.}
    \label{fig:BendingConfig}
\end{figure}

A rectangular-shaped cantilever with a rounded base geometry was fabricated using a FIB in a FEI Quanta 3D dual beam scanning electron microscope (SEM). The sides of the bream cross section were measured as $a_z=a_y=5.0\pm0.1\,\mathrm{\upmu m}$, and the length of the beam was $L=20.6 \pm0.1$\,$\mathrm{\upmu}$m. In order to avoid stress concentrations at sharp edges, circular geometry was used at the base of the cantilever with a diameter of $d=8.0\pm0.1\,\mathrm{\upmu}$m (Fig.~\ref{fig:BendingConfig}). During the loading, the indenter's bending tool was in contact with the cantilever along a straight line parallel to the cantilever's $z$ direction, and was $2.1 \pm 0.2\,\upmu$m far from the free end of the cantilever which resulted in a moment arm length of $l = 18.5 \pm0.3$\,$\mathrm{\upmu}$m. During microsample preparation, the final FIB polishing step was operated at 30~kV and a Ga$^{+}$ ion current of 500~pA was used, which resulted in less than $1^\circ$ differences between the parallel sides of the cantilever and a minimal Ga$^+$ implantation. It is noted, that in total 3 microcantilevers were prepared with identical geometry and orientation, and since all of them exhibited equivalent plastic behaviour, in the paper we focus on the results obtained for a single measurement.

\subsubsection{Micromechanical deformation}

The microcantilever bending experiments were performed using a custom-made \emph{in situ} microdeformation stage \cite{hegyi2017micron, ispanovity2022dislocation}. The hook-shaped bending tool was made from a tungsten needle tip by FIB machining. The deformation tool was attached to a spring with a spring constant of $1.7$~mN/$\upmu$m that was used to determine the acting force. During bending, the spring was moved with a constant velocity of $\pm5\mathrm{\:nm/s}$.

\subsubsection{EBSD/HR-EBSD}
\label{sec:hrebsd}

The orientation of the sample was determined using EBSD. An Edax Hikari camera was used to record the patterns with 1 $\times$ 1 binning, using an electron beam of 20 kV, 4 nA. The measurement confirmed that the sample consists of a single grain and the orientation was found to be $5.2^\circ$ inclined w.r.t.~the preferred $\langle100\rangle \{100\}$ orientation, as indicated in Fig.~\ref{fig:BendingConfig}b (see also Suppl.~Fig.~\ref{fig:C3-EBSD-orient}). In order to characterise the plastic zone in the deformed cantilevers, \emph{in situ} HR-EBSD was utilized on the FIB-machined surface of the cantilever. HR-EBSD is a surface technique where the diffraction patterns collected on the specimen are related to a reference pattern through image cross-correlation \cite{Wilkinson2006, Wilkinson2009}. The special geometry of the indentation device allowed to perform some of the HR-EBSD mappings \emph{in situ}, while the sample was under load. Therefore, during such mappings (which took approx.\ 25 minutes), the motion of the hook was paused. In other cases, HR-EBSD maps were recorded after releasing the sample, creating a unique opportunity to compare the HR-EBSD maps obtained under load and after releasing the sample to a relaxed stress state. HR-EBSD evaluation was performed using BLGVantage CrossCourt Rapide v.4.5.0.

The characterization of the microstructure is given using the elements of the stress tensor, the net GND density and the GND density tensor, defined by Nye as:
\begin{eqnarray}\label{eq:01}
\alpha_{ij} = \sum_t b_i^t l_j^t\rho^t, \label{eqn:alpha}
\end{eqnarray}
where index $t$ represents a slip system, $\rho^t$ is the density on that system and $\bm b^t$ and $\bm l^t$ is the Burgers vector and line direction of dislocations, respectively. From the elements of the distortion tensor $\beta_{ij} = \partial_j u_i$ with $\bm u$ being the displacement field, accessible by HR-EBSD, three components of the tensor $\alpha$ follow as \cite{Wilkinson2009}:
\begin{eqnarray}\label{eq:ai3}
\alpha_{i3} = \partial_1 \beta_{i2} - \partial_2 \beta_{i1}, \quad i=1,2,3.
\end{eqnarray}
According to Eq.~(\ref{eqn:alpha}) the $\alpha_{i3}$ components are a sum in which all types of dislocations not parallel with the $xy$ plane have a contribution. This means that, in principle, from the $\alpha_{i3}$ values themselves one cannot asses the amount of dislocations with different type $t$, that is $\rho^t$. If, however, there is a dominant type of BVs in the population, then the sum in Eq.~(\ref{eqn:alpha}) will be dominated by that term and the $B_i = \alpha_{i3}$ vector is nearly parallel with $b^t_i$ (for the geometrical interpretetaion of the meaning of the signs of the Nye tensor components see Fig.~\ref{fig:BendingConfig}). Exactly this was found earlier for Cu micropillars \cite{zoller2021microstructure} as well as for bulk Cu samples \cite{lipcsei2022statistical}. Here we thus apply the same method described in \cite{zoller2021microstructure} to determine the dominant BV type at every measurement point of the HR-EBSD.

\subsubsection{TEM}
\label{sec:methods_tem}

Bright field (BF) imaging was performed on a Cs-corrected (S)TEM Themis TEM with 200 kV electron beam. One lamella with a surface normal $\vec{n} \parallel [001]$ was removed from the sample after the deformation experiment using a Thermo Fisher Scios 2 Dual Beam microscope. In order to determine the proportion of dislocations with different BVs in the sample, 5 different tilt positions were used. In the [001] zone condition all the dislocations were visible, whereas under (200), (220), (020) and $(2\bar{2}0)$ two-beam conditions dislocations with slip directions 1+2, 5, 3+4 and 6 (cf.~Fig.~\ref{fig:slip_systems}b), respectively, were not seen. This allowed us to validate the HR-EBSD measurements, however, the TEM foil could only be extracted from somewhat beneath the surface.

\subsection{Simulation}
\subsubsection{Dislocation density based continuum model}
\label{sec:CDD}

The dislocation motion in the 3D space can be described for homogenised ensembles of curved dislocation lines by the kinematically closed Continuum Dislocation Dynamics (CDD) formulation derived by Hochrainer \emph{et al.}\ using a higher-dimensional space preserving information about the dislocation orientations \cite{hochrainer2007three,hochrainer.2014,hochrainer2015multipole}.
The CDD formulation used here is based on the framework in \cite{schulz.2019} and incorporates two coupled problems. In the external problem the stress field is calculated for a given plastic state and in the internal problem the microstructure evolution is derived for a given stress field yielding plastic deformation. These two problems are thus coupled via the plastic slip, which is assumed to be the sole result of collective dislocation motion.

The microstructure is described by different dislocation densities and a curvature density on the individual slip systems, whereby following \cite{sudmanns.2020} we distinguish between mobile dislocation density, yielding plastic deformation, and immobile dislocation network density, contributing to the material hardening. The mobile dislocation density is again additively decomposed into its statistically stored dislocation (SSD) and geometrically necessary dislocation (GND) parts, whereby the screw and edge character of the GND density is divided.
A visualisation of the degrees of freedom of the CDD formulation used is given in \cite{zoller2021microstructure}.

The microstructure evolution also incorporates activities of dislocation sources covered by the homogenised model introduced in \cite{zoller.2020}, dislocation multiplication mechanisms including cross-slip and glissile reactions according to \cite{sudmanns.2019}, dislocation network formation due to Lomer reactions and annihilation processes due to collinear reactions given in \cite{sudmanns.2020}.
Depending on the local stresses and dislocation configuration, the Lomer junctions can unzip again, whereby the link lengths of dislocation segments connected to Lomer junction, here called Lomer arms, influence the critical stress for unzipping \cite{rodney.1999,shin.2001}.
Following \cite{zoller2021microstructure}, the link length distribution of these stabilised Lomer arms are modelled by a Rayleigh distribution with an expected value equal to to average link length, which is assumed to scale with the averaged dislocation distance.
The constitutive law for the dislocation velocity takes internal stresses into account including short-range dislocation stress interactions represented by a back-stress term given in \cite{groma.2003,schmitt.2015} and eigenstresses of the dislocations according to \cite{schulz.2014}. The hindrance of dislocation motion due to interaction with forest dislocations from other slip systems within the averaging volume are considered by a yield stress term based on \cite{franciosi.1980}. The resulting plastic slip on the individual slip systems is described by the Orowan equation \cite{orowan.1934}.
The formulation used focuses on small deformations and the stress-strain relation describes physical linearity using the cubic symmetry of the elasticity tensor. The Cauchy stress tensor inserted in the static momentum balance ensures the macroscopic equilibrium condition.

\subsubsection{Numerical implementation}
The CDD formulation used was implemented into a two-scale numerical framework based on \cite{schulz.2019} using an own customised version of the parallel finite element software M++ \cite{wieners.2005,wieners.2010}. A finite element approach is applied with hexahedral elements and linear ansatz functions for the displacements in the external problem as well as an implicit Runge-Kutta discontinous Galerkin scheme with full upwind flux and constant ansatz functions for the dislocation and curvature densities in the internal problem. Simplifying but found to be numerically efficient, both scales are meshed with the same spatial resolution. The time is discretised by an implicit midpoint rule with a fixed time step.

\subsubsection{Simulation parameters}
\label{sec:CDD_parameter}

The main parameters of the CDD formulation are summarized in Table \ref{tab:01}.

\begin{table}[h]
\caption{Main parameters of the CDD formulation}\label{tab:01}
\begin{tabularx}{0.49\textwidth}{|l|X|X|} 
 \hline
 Parameters & Values & References \\ \hline
 Anisotropic & $C_{1111}=168\,\mathrm{GPa}$ & \cite{ledbetter.1974,rosler.2019} used in \\ 
elastic & $C_{1122}=121\,\mathrm{GPa}$ & Eq.~(5) in \cite{schulz.2019}  \\ 
constants  & $C_{2323}=75\,\mathrm{GPa}$ &  \\ \hline
Isotropic    &  $\mu=40\,\mathrm{GPa}$ &  Eqs.~(15) \& (17)  \\
elastic    & $\nu=0.367$\  & in \cite{schulz.2019}, derived  \\
constants    &   &  following \cite{date.1969} \\ \hline
BV length    &  $b=0.254\,\mathrm{nm}$ &  \cite{davey.1925} \\\hline
Drag coefficient    & $B=5\times10^{-5}\,\mathrm{sPa}$ &  \cite{kubin.1992}, \cite{schulz.2019} Eq.~(14) \\\hline
Interaction  &  $a_{\mathrm{self}}=0.3$ &  \cite{Akhondzadeh.2020}, \cite{schulz.2019} Eq.~(17)\\
matrix    & $a_{\mathrm{copl}}=0.152$  &  \\
    & $a_{\mathrm{Hirth}}=0.083$  &   \\
    & $a_{\mathrm{Lomer}}=0.326$  &   \\
    & $a_{\mathrm{gliss}}=0.661$  &   \\
    & $a_{\mathrm{coll}}=0.578$  &   \\ \hline
Backstress  &  $D=\frac{3.29}{2\pi^2(1-\nu)}$ &  \cite{schmitt.2015}, \cite{schulz.2019} Eq.~(15) \\\hline
Cross-slip    &  $\beta=10^5$ &  \cite{weygand.2002} \\
probability    &   &   \\\hline
Activation    & $V_\mathrm{act}=300\,b^3$  &  \cite{bonneville.1988} \\
volume    &   &   \\\hline
Initial stress &  $\tau_\mathrm{III}=28\,\mathrm{MPa}$  &  \cite{kubin.1992}, \cite{sudmanns.2019} Eq.~(20)\\
for Stage III    &   &   \\\hline
Dislocation    &  $c_\mathrm{gliss}=0.06$ &  \cite{zoller2023classification}, \cite{sudmanns.2020} Eq.~(13)\\
reaction    &  $c_\mathrm{Lomer}=0.09$ &   \\
coefficients    &  $c_\mathrm{coll}=0.003$ &   \\\hline
\end{tabularx}
\end{table}

It should be noted that due to the lack of data for the dislocation reaction coefficients in the case of copper under bending, the exact choice of the dislocation reaction constants is uncertain and the values were assumed to be equal to the modelled gold under torsion in \cite{zoller2023classification}.
The initial microstructure is assumed to consist of a total dislocation density of $\rho^\mathrm{Tot}=7.2 \times 10^{13}\,\mathrm m^{-2}$, which is homogeneously distributed on all slip systems. Since the degree of connectivity of the initial dislocation network is unknown, it is assumed that the initial total dislocation density consist half of mobile and half of network dislocation density.
The applied external load is displacement-controlled at a rate of $|\dot{u}_\mathrm D|=0.1$ $\mathrm{ms}^{-1}$.

\section{Results}
\label{sec:results}

\subsection{Microstructure characterisation}
\label{sec:microstructure_characterisation}

\begin{figure}[!h]
    \centering
    \includegraphics[width=0.45\textwidth]{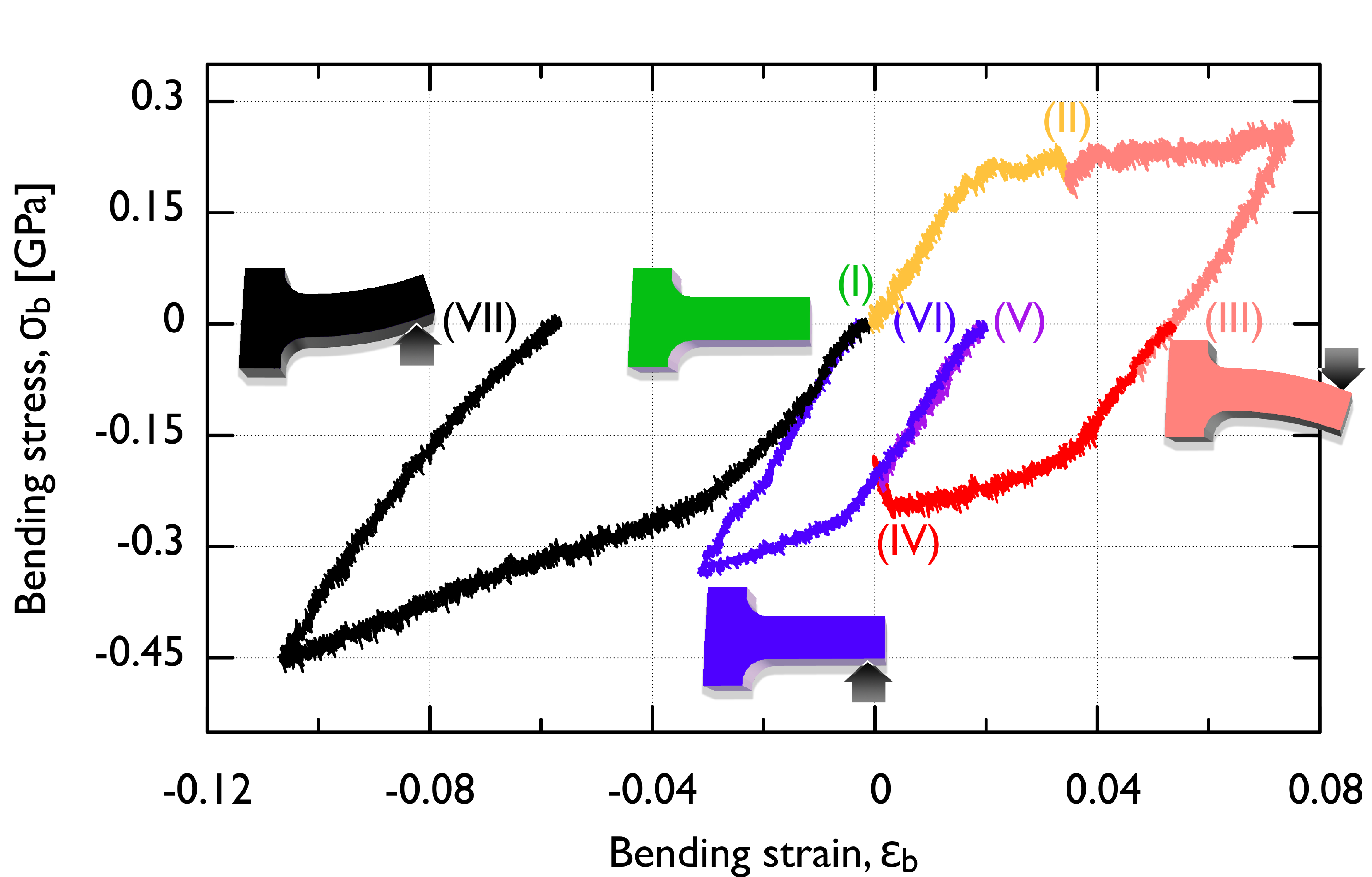}
    \caption{Bending stress-bending strain curve of the microcantilever. The bending experiment was performed in 6 steps, this is indicated by the different colours.  HR-EBSD scans were conducted at the initial state and after each step, as indicated by the roman numerals (I)--(VII). The four sketches illustrating the state of the cantilever are to help the interpretation of the figure and the arrows represent the bending direction prior to the actual measurement. For the corresponding force-displacement curve, see Suppl.~Fig.~\ref{fig:C3-ld2}.}
    \label{fig:C3-ld}
\end{figure}

The cantilever was first bent in one direction (along the $y$ axis in the negative direction), followed by reversed loading up to a similar extent. The measured plastic response of the sample is shown in Fig.~\ref{fig:C3-ld}, where stresses and strains are normalised according to the theory of elastic bending and the corresponding literature \cite{motz2005mechanical,demir2010mechanical,Husser.2017}, as:
\begin{equation}
    \sigma_\mathrm b = \frac{16F}{a^2}
\end{equation}
and for the bending strain the ratio of the vertical displacement $u_\mathrm D$ and the moment arm $l$ is considered:
\begin{equation}
    \varepsilon_\mathrm b = u_\mathrm D/l.
\end{equation}
(For the corresponding force-displacement curve see Suppl.~Fig.~\ref{fig:C3-ld2}.) During the first bending, a clear strain hardening can be observed until the maximum plastic bending strain (measured after elastic unloading) of $\varepsilon_\mathrm{p, b} \approx 5.5 \%$ is reached. Upon reversed loading, the Bauschinger effect is clearly seen as well as a significantly stronger strain hardening compared to the initial direction. To reach the terminal plastic strain of $\varepsilon_\mathrm{p,b} \approx -6 \%$ in the reverse direction requires an approx.\ 70\% larger bending stress compared to the initial direction.

\begin{figure*}[!h]
    \centering
    \includegraphics[width=\textwidth]{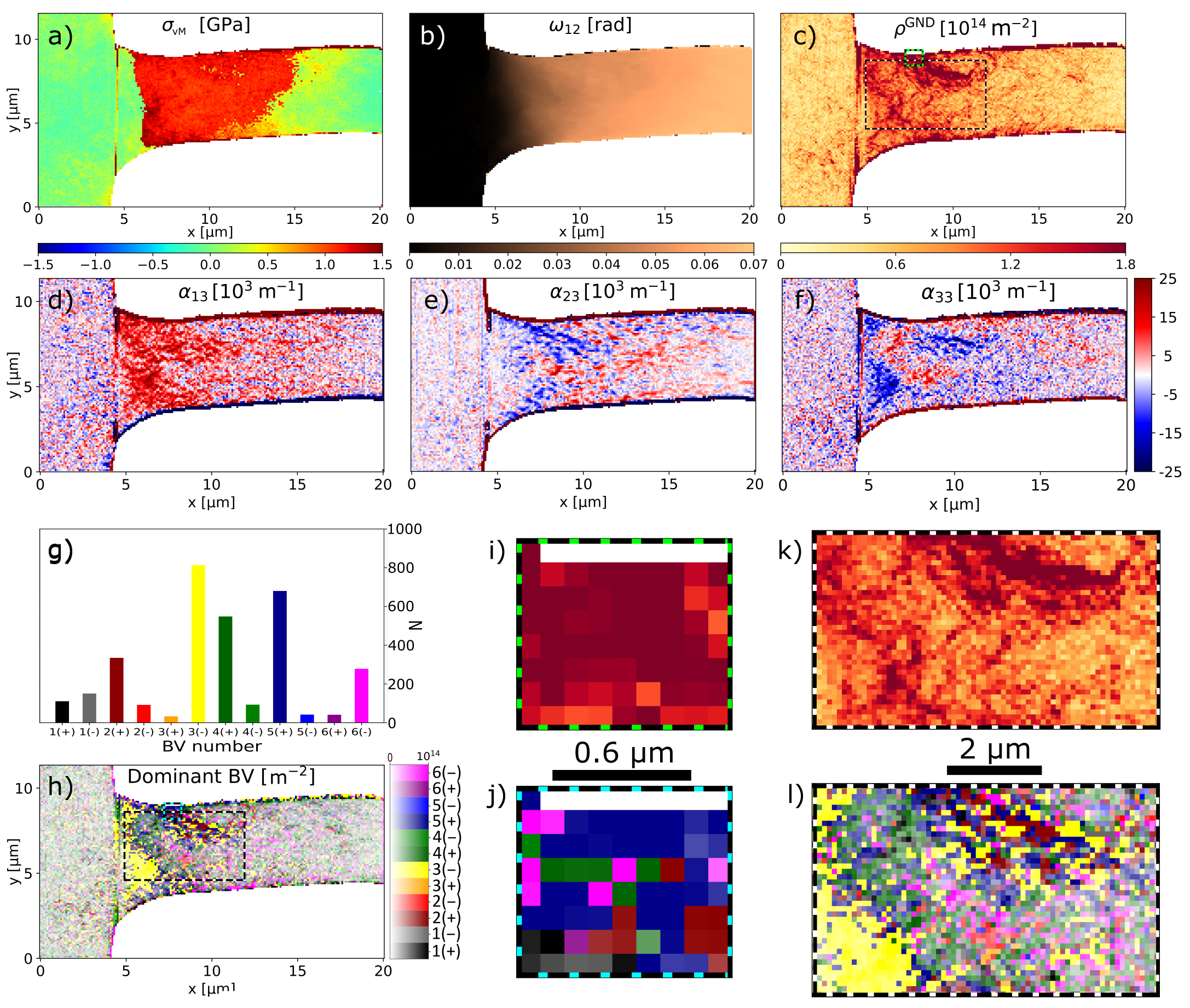}
    \caption{Results of the HR-EBSD microstructural analysis for stage (VII). (a)-(c): Components of the von Mises stress tensor $\sigma_\text{vM}$, rotation tensor $\omega_{12}$ and the GND density $\rho^\text{GND}$. (d)-(f) Components $\alpha_{i3}$ of the Nye dislocation density tensor. (g): Histogram of the number of dislocations $N$ (obtained as $N = \rho^\text{GND} d^2$, where $d$ is the step size of the EBSD scan) for different BV types. (h): Distribution of the dominant BVs, colours indicate the BV type, and the darkness corresponds to the GND density, as shown by the 2D colourbar. The rectangles with dashed and solid lines indicate the areas for the BV analysis and the TEM foil, respectively. (i)-(l): Insets showing the magnified views of panels (c) and (h) of the areas of the BV analysis and TEM foil.}
    \label{fig:hrebsd}
\end{figure*}

HR-EBSD measurements were performed seven times during the course of the experiment at stages denoted by roman numerals (I)--(VII). Measurements (II) and (IV) were carried out \emph{in situ} under load, which explains the slight drops in the stress due to relaxation phenomena. The other five scans were performed at zero applied stress, that is, after releasing the sample. After each HR-EBSD scan the sample surface gets slightly contaminated that causes the deterioration of the EBSD pattern quality of subsequent scans at the same area. To eliminate the negative consequences of this effect a few tens of nm thick layer was removed by FIB from the sample surface after each scan with the same setting as was applied in the final polishing step during the cantilever preparation. To perform this, the sample had to be dismounted from the device after every scan and then had to be re-mounted again (the colouring of the different parts of the stress-strain curve is meant to reflect this detail of the experiment). A slight misorientation of the sample is unavoidable during this process, which explains the slight drop in the yield stress after measurement (VI). This artefact, however, is not expected to play any role in the general trends of the behaviour to be discussed in the following. It is noted that between measurements (IV) and (V) only elastic unloading and the weak FIB-polishing discussed above took place which allowed us to check the influence of these two processes on the microstructure.

Figure \ref{fig:hrebsd} presents the HR-EBSD based evaluation of the microstructure for the final stage (VII). In panel (a) the von Mises stress $\sigma_\mathrm{vM}$ calculated from the components of the stress tensor (the elements of which are accessible with HR-EBSD, as mentioned in Sec.~\ref{sec:hrebsd}) is plotted. 

The GNDs are also required to accommodate the lattice rotation due to plastic bending along the $z$ axis that is shown in panel (b) (note that the orientation changes rather smoothly, no sharp jumps can be observed that would correspond to simple low angle grain boundaries). Indeed, the total GND density [panel (c)] shows a well-developed inhomogeneous distribution of dislocations resembling a cellular structure also seen in bulk samples \cite{kalacska2017comparison, lipcsei2022statistical}. The components of the Nye tensor $\alpha_{i3}$ show extended regions with non-zero net BV [panels (d-f)].  Panel (h) shows the result of the BV analysis introduced in Sec.~\ref{sec:hrebsd}. As seen, large areas characterised by a single dominant BV (e.g., at the base of the cantilever almost exclusively dislocations of type 3 are found) are alternating with regions with more complex microstructure. The figure also differentiates between different signs of the BVs. As seen visually and also supported by the histogram in panel (g), for every BV type one direction is strongly preferred, that is, the microstructure is highly polarised due to the lattice rotations (that is, strain gradients), as expected. In order to eliminate measurement artefacts at the sample edges, the histogram is calculated for the central area of panel (h) (denoted with dashed line and also shown in a magnified view in panels (k) and (l)) where most of the GNDs are accumulated.

\subsection{TEM experiments}

\begin{figure}[!h]
    \centering
    \includegraphics[width=0.49\textwidth]{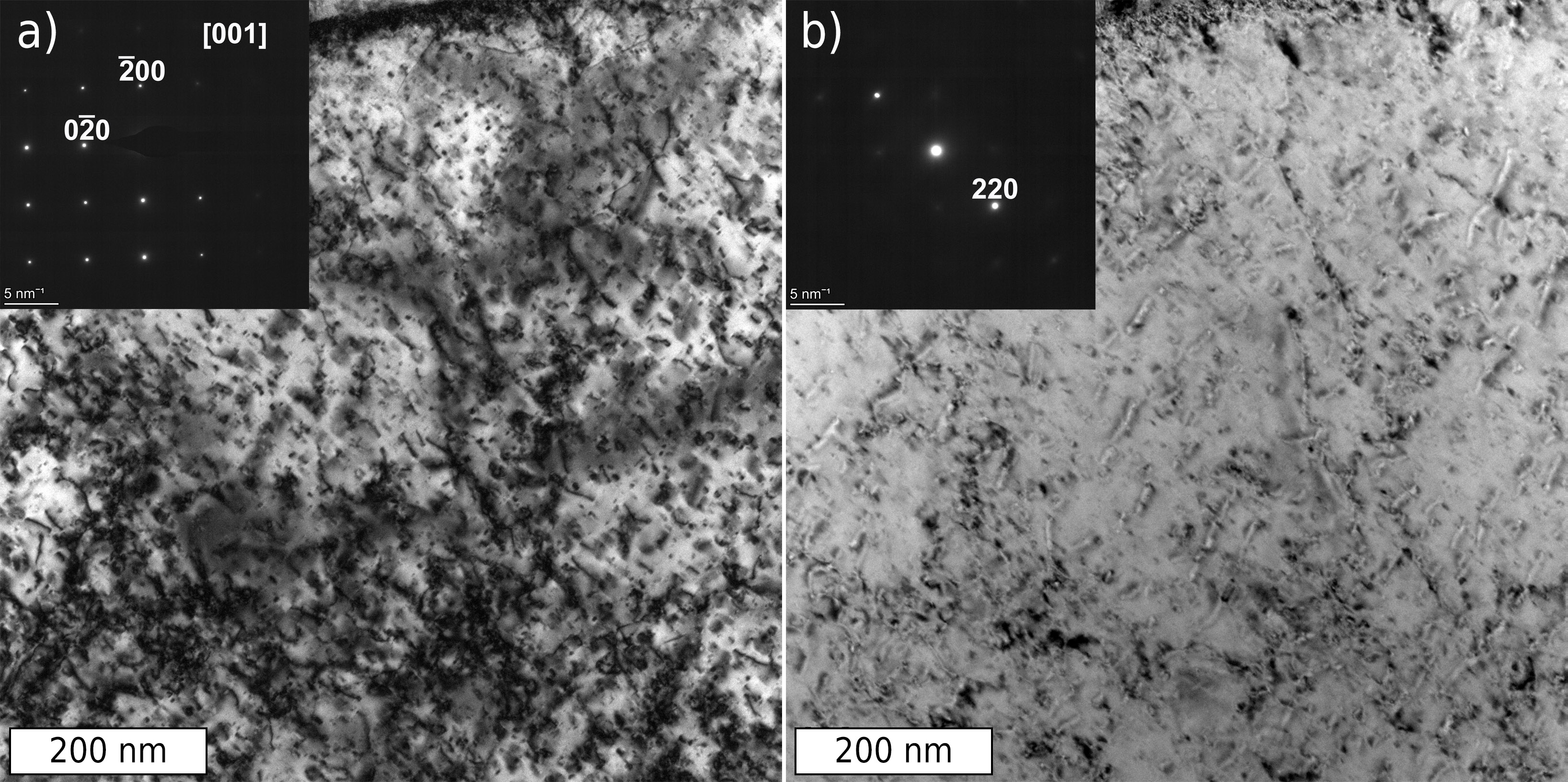}
    \caption{TEM micrographs showing the dislocation content of the region indicated with solid line in panels c) and h) of Fig.~\ref{fig:hrebsd}. a) BF in zone [001], b) BF in (220) two-beam condition.}
    \label{fig:tem}
\end{figure}

BF images made from the [001] zone axis and (220) two-beam conditions are shown in Fig.~\ref{fig:tem}. For the other BF images taken from the same area see Suppl.~Fig.~\ref{fig:tem_supp}. In Fig.~\ref{fig:tem}a a well-developed cellular dislocation structure is seen with a large number of dislocations even in the cell interiors. Similar conclusions can be drawn from the images taken at the (200) and (020) two-beam conditions (Suppl.~Figs.~\ref{fig:tem_supp}a and \ref{fig:tem_supp}b). However, under $(2\bar{2}0)$ and (220) conditions, the contrast of most dislocations disappear (as seen in Fig.~\ref{fig:tem}b and Suppl.~Fig.~\ref{fig:tem_supp}c). Based on these observations and the discussion of Sec.~\ref{sec:methods_tem} it can be concluded that the majority of the dislocations found in the lamella had $[\bar{1}10]$ or $[110]$ type BVs, that is, of type 5 and 6 or, equivalently, Group II. This finding is in good agreement with the HR-EBSD evaluation, as in Fig.~\ref{fig:hrebsd}l 80 \%  of the dislocations were found to fall into this group.

\subsection{Evolution of the microstructure}

\begin{figure}[!ht]
    \centering
    \includegraphics[width=0.36\textwidth]{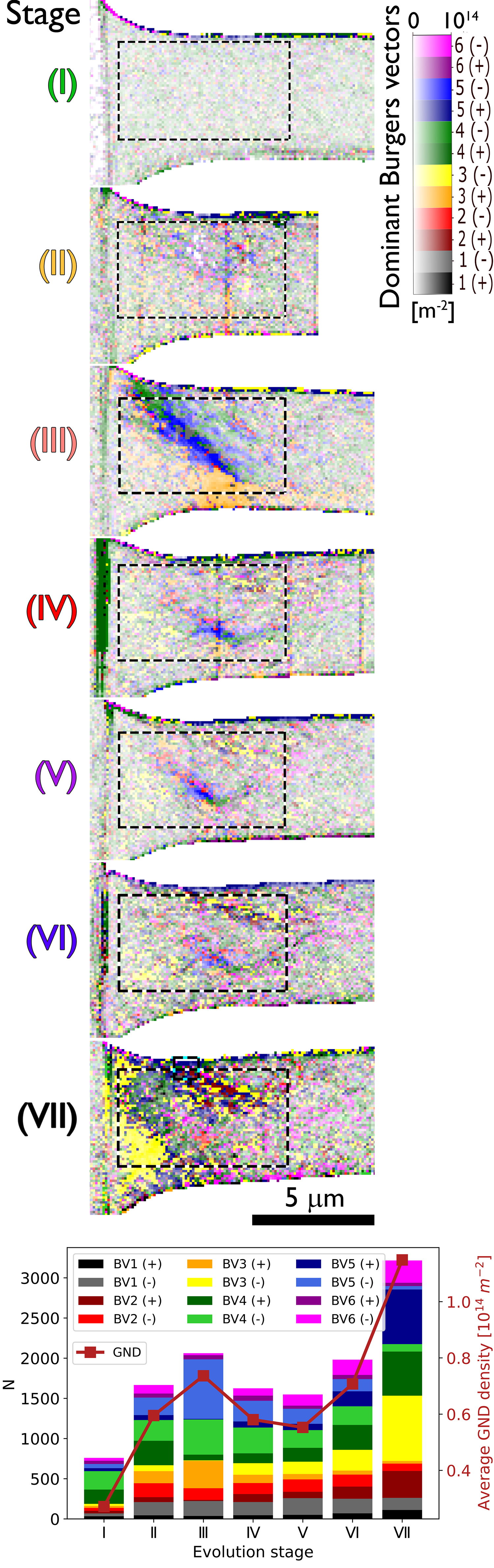}
    \put(-200,590){\sffamily{a)}}
    \put(-195,115){\sffamily{b)}}
    \caption{Evolution of the dislocation microstructure. (a) Distribution of the dominant BVs at subsequent deformation steps. Black dotted lines mark the areas where BV statistics were collected. (b) The number of dislocations $N$ of various BV types as a function of the deformation stage.}
    \label{fig:evolution_exp}
\end{figure}

Figure \ref{fig:evolution_exp} is concerned with the evolution of the dislocation microstructure during bi-directional bending. Panel (a) shows the GND densities and the dominant BV types for stages (I)-(VII) shown in Fig.~\ref{fig:C3-ld} (for the corresponding components of the Nye tensor and the GND densities see Suppl.~Figs.~\ref{fig:alpha_evolution} and \ref{fig:GND_evolution}). As seen, in the initial state no specific dislocation structure can be recognized, the measured densities of BV type 4 are due to the noisy background. As stress starts to increase at stages (II) and (III) a dislocation pile-up gradually builds up with a majority of dislocations with BV type 5 in the central region. Note, that the structure is strongly polarised, so the thin region with BV type 5 extending in the $[\Bar{1}10]$ direction is practically a low-angle grain boundary separating the base of the beam from the bent part. Although the BV type 5 also points in the $[\Bar{1}10]$ direction, the slip planes are not perpendicular to the $xy$ plane and the structure exhibits a width of approx.~1 $\upmu$m, so this structure is clearly not a simple pile-up in a single glide-plane but the constituent dislocations are rather expected to have originated from sources outside the plane of the pile-up.

Upon reversed loading this structure gradually dissolves during stages (IV)-(V), first close to the top/bottom surface, and then a lot of additional GNDs accumulate throughout the cross section of the cantilever during stages (VI)-(VII). Accordingly, the total number of GNDs initially start to drop until stage (V) and then increase to a much higher value compared to stage (III) as seen in panel (b) of Fig.~\ref{fig:evolution_exp}.  In the meantime the polarisation of the microstructure reverses as all present BVs change their signs. The final dislocation structure is rather complex, with the dominant BV alternating quite quickly in space. This complexity suggests a large number of dislocation interactions taking place between different slip systems during reversed bending. This and the large increase in the GND density explains the high strain hardening during this phase and the enhanced bending stress (450 MPa compared to 200 MPa of the initial bending). The results are in a very good agreement with the general predictions of Sec.~\ref{sec:system_analysis} as the great majority of dislocations fall into Groups I and II.

As it was mentioned earlier, between stages (IV) and (V) only elastic unloading and a slight FIB polishing were performed. According to Fig.~\ref{fig:evolution_exp} the microstructures in the two cases show identical features in BV distributions and GND content, so we conclude, that elastic unloading and the FIB polishing have negligible effects on the microstructure observed by HR-EBSD.

To explore the origin of the GNDs, high quality SEM
images were made on the top and bottom surfaces of cantilevers that were fabricated identically to the \emph{in situ} sample. This was required because the re-deposition due to the consecutive FIB polishing steps (performed between each deformation stage because of the degradation of the EBSD pattern quality as mentioned before) masked these fine patters and it was not possible to record them afterwards on the original sample. We assume that identical slip traces would have been found if the FIB polishing was not essential to keep the EBSD mapping quality as high as possible. On the top side a characteristic checkered pattern was visible that corresponds to slip traces of dislocations gliding on different $\{111\}$ planes (Suppl.~Fig.~\ref{fig:SEM-pattern}).

\subsection{Simulation}

\begin{figure}[!h]
    \centering

    \hspace*{-1.0cm}
    \includegraphics[width=0.41\textwidth]{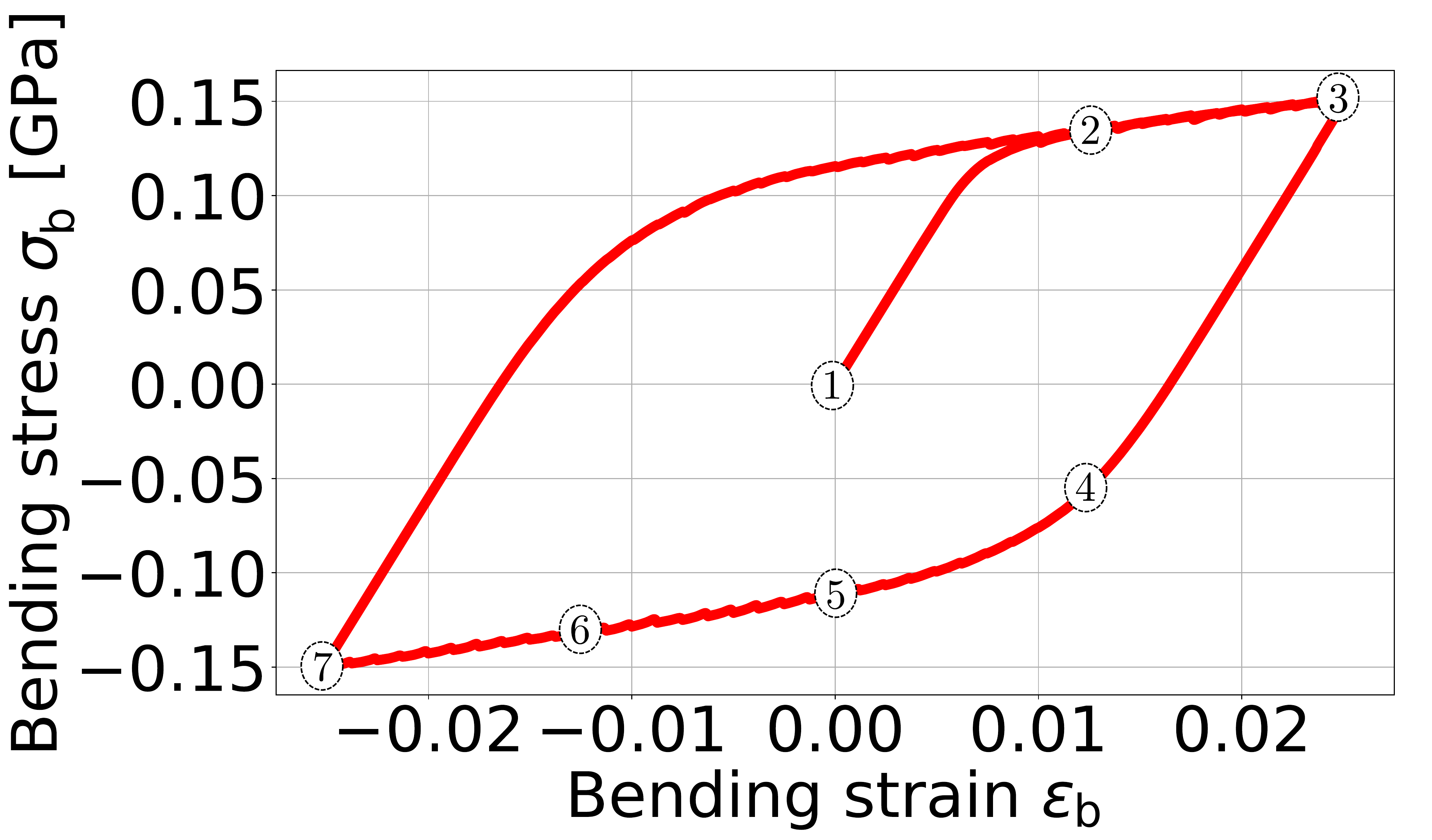}
    \put(-190,125){\sffamily{a)}}

     \hspace*{0.3cm}
     \includegraphics[width=0.45\textwidth]{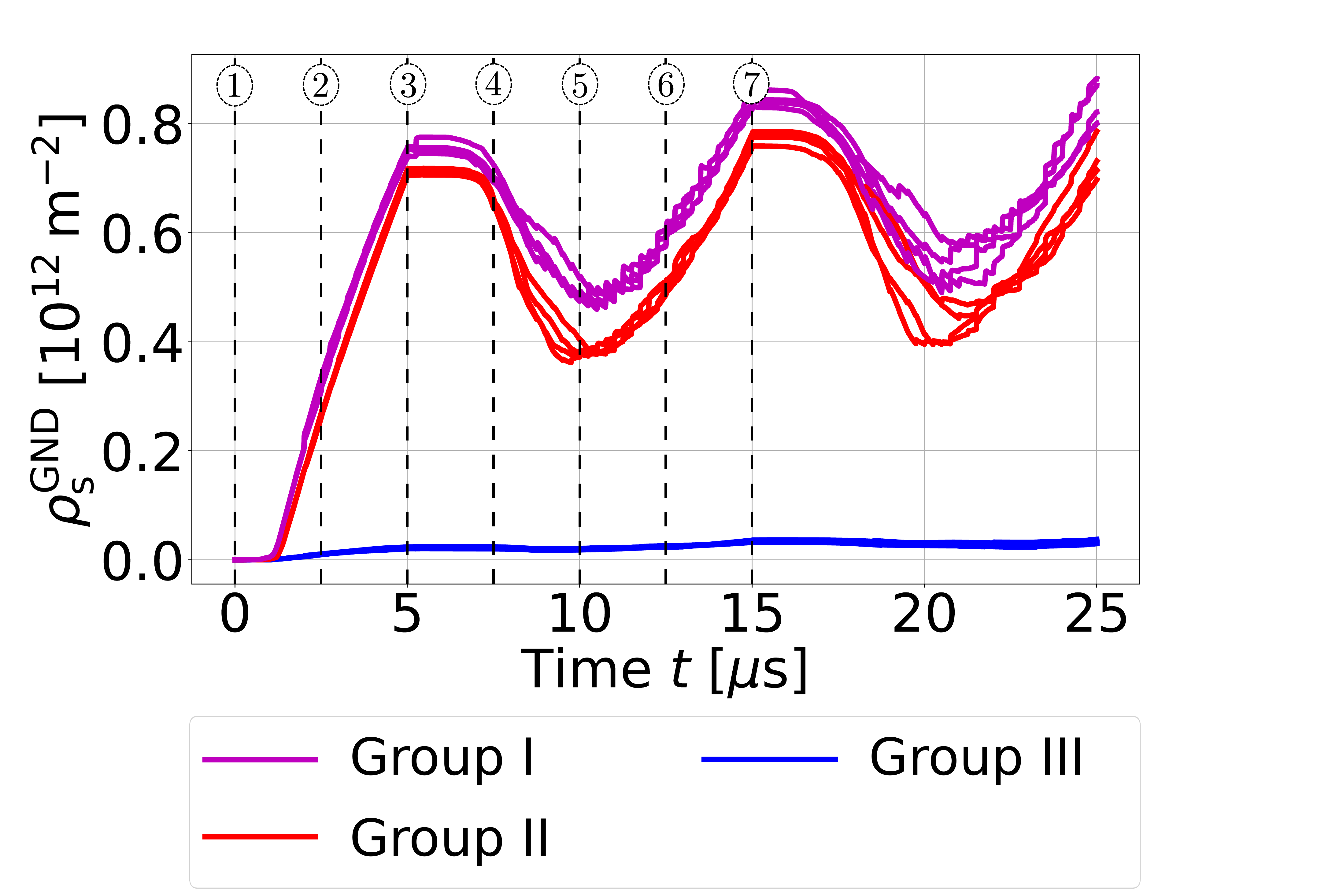}
     \put(-220,150){\sffamily{b)}}

    \caption{CDD simulation of bi-directional bending. (a): Cycle of bending moment based on the displacement driven loading and (b): resulting GND density evolution for the 12 different slip systems grouped according to Sec.~\ref{sec:system_analysis}.}
    
    \label{fig:cycleloading}
\end{figure}

To get deeper insights into involved slip systems during the cantilever bending, the experiments are complemented by CDD simulations. Considering the simulation parameters as described in Sec.~\ref{sec:CDD_parameter}, we analyse the GND density evolution with respect to a given loading over the system. For a single loading cycle as shown in Fig.~\ref{fig:cycleloading} (top), the resulting GND density evolution is given in Fig.~\ref{fig:cycleloading} (bottom) for all 12 slip systems. It is noted, that due to computational limitations the bending amplitude in the simulations is approximately three times smaller than in the experiments. As such, in this section we aim mostly at a qualitative, rather than quantitative comparison with the experimental results.
\begin{figure}[b]
    \centering
    \includegraphics[width=0.495\textwidth]{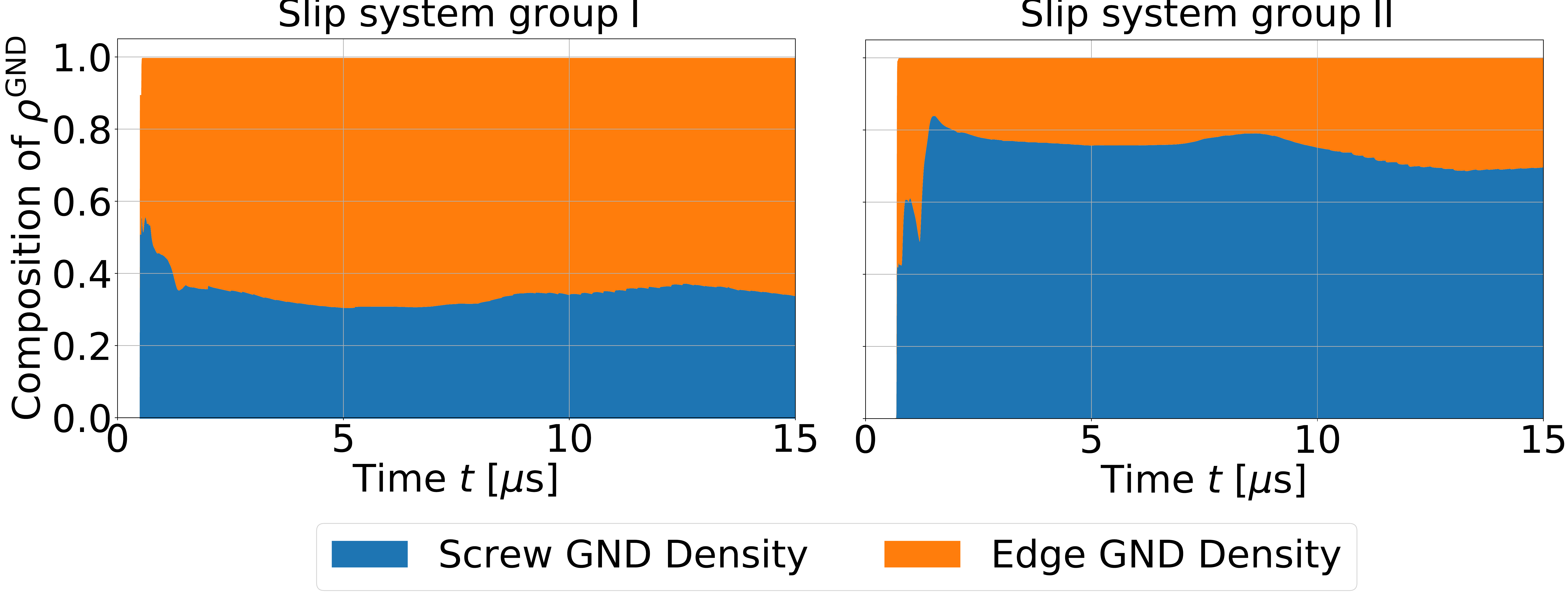}
    \caption{Comparison of the composition of the screw and edge dislocation density for the active slip system groups.}
    \label{fig:scre_edge}
\end{figure}

Applying the slip system denomination as given in Fig.~\ref{fig:slip_systems}, it can be observed that the slip system activity can be distinguished between the three different slip system groups introduced in Sec.~\ref{sec:system_analysis}. Group III is found to be almost inactive. Based on the system analysis, this group comprises slip systems that see mainly shear stresses during elastic loading while Groups I and II contain systems with predominantly normal or solely normal stress contributions to the resolved shear stress on the respective systems. The GND density evolution reflects the loading and unloading paths by a wavy course. This occurs very similarly for the two active slip groups.

Looking into the loading for a half cycle (up to the maximum negative loading, shown as stages 1-7), it can be seen how the change of the loading direction results in a highly nonlinear GND evolution showing slightly higher values for Group I compared to Group II.

Looking closer into the two active slip system groups, the character of the dislocation content is analysed. Fig.~\ref{fig:scre_edge} depicts the composition of screw and edge components for the active slip system groups. The comparison shows that the composition is different: for Group I the amount of screw dislocations is lower compared to the amount of edge dislocations and their ratio is approximately constant all over the simulation time.

\begin{figure}[h]
    \centering


      

   \includegraphics[width=0.49\textwidth]{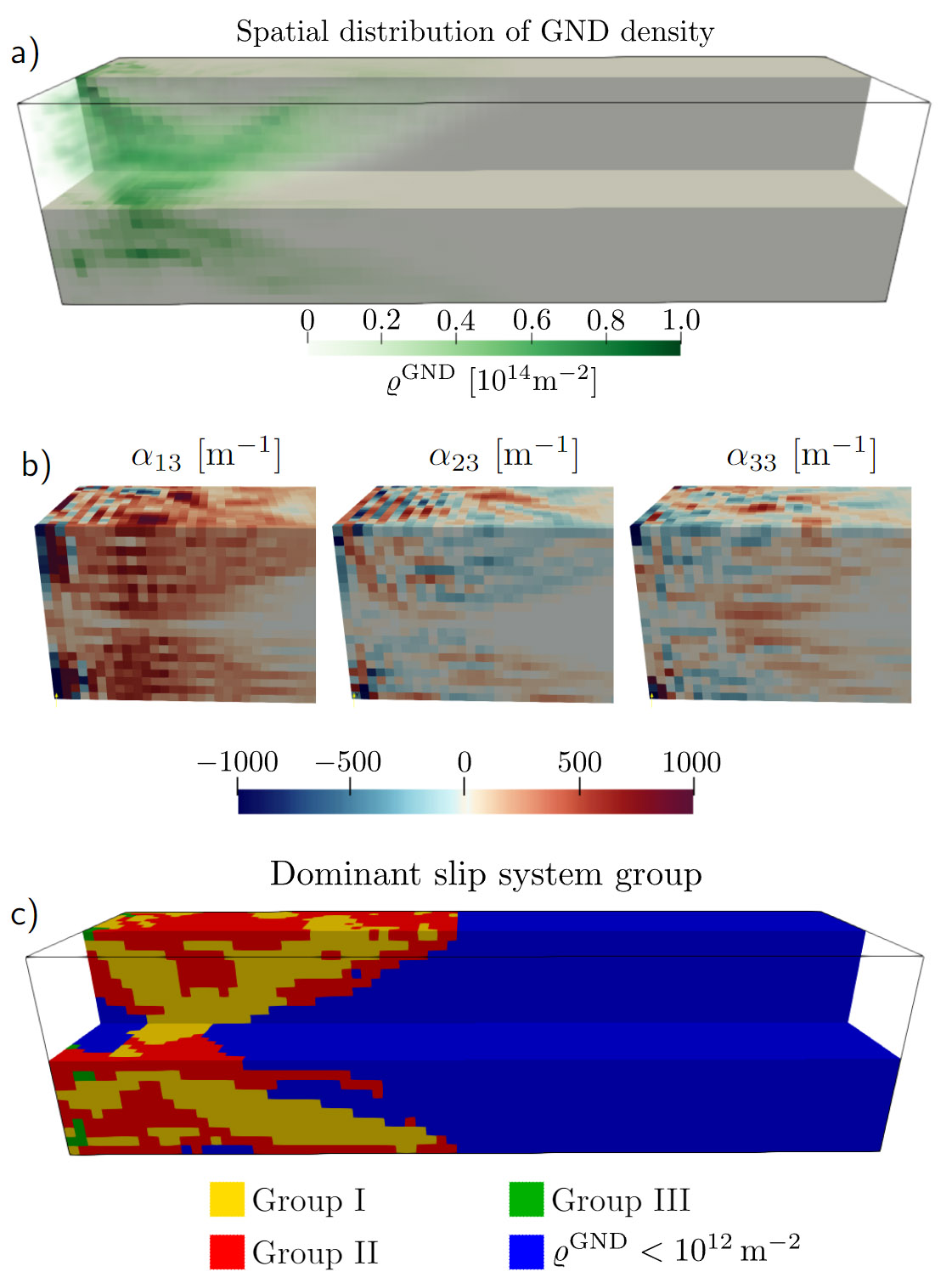}
    
    \caption{Distribution of GNDs at stage 7 of the CDD simulations. (a): The spatial distribution of the GND density, (b): contour plots of the tensor components $\alpha_{i3}$ and (c): the dominant slip system groups over the cantilever beam.}
    \label{fig:contourCantilever}
\end{figure}
For Group I, on the other hand, the amount of screw dislocation density is higher compared to the edge dislocation density. This difference is probably due to the effect mentioned in Sec.~\ref{sec:system_analysis}, that is, dislocations in Group I have pure screw character in the neutral zone which leads to much higher cross-slip probabilities. This process can at the same time explain the lack of screw dislocations in Group I and the excess in Group II.

The contour plot of the whole cantilever in Fig.~\ref{fig:contourCantilever}a reveals the location of the slip activity and the occurring dislocation density pileups. The spatial distribution of GND density emerges mainly in the region close to the clamped edge due to the stress concentration. At the maximum loading, 95\% of the GND density can be found in the region $x/l\leq 0.4$. Averaged over cross section, the maximum averaged values can be found at $x/l\approx 0.1$. High density values are found close to the neutral axis. In direct proximity to the clamped edge as well as in the transition zone to almost GND-free regions there are distinctive pileups found along the diagonal in the $xy$ plane. For the regions of high activity, the tensor components $\alpha_{i3}$ are given in Fig.~\ref{fig:contourCantilever}b.
Focusing on the composition of the dislocation density with respect to the different slip system groups, we observe a complex interplay of the two active groups, see Fig.~\ref{fig:contourCantilever}c. Here, elements with densities lower then $\rho^\mathrm{GND}<10^{12}\, \mathrm m^{-2}$ are neglected in the evaluation and are coloured in blue.

\begin{figure*}[h]
    \centering
    \includegraphics[width=0.95\textwidth]{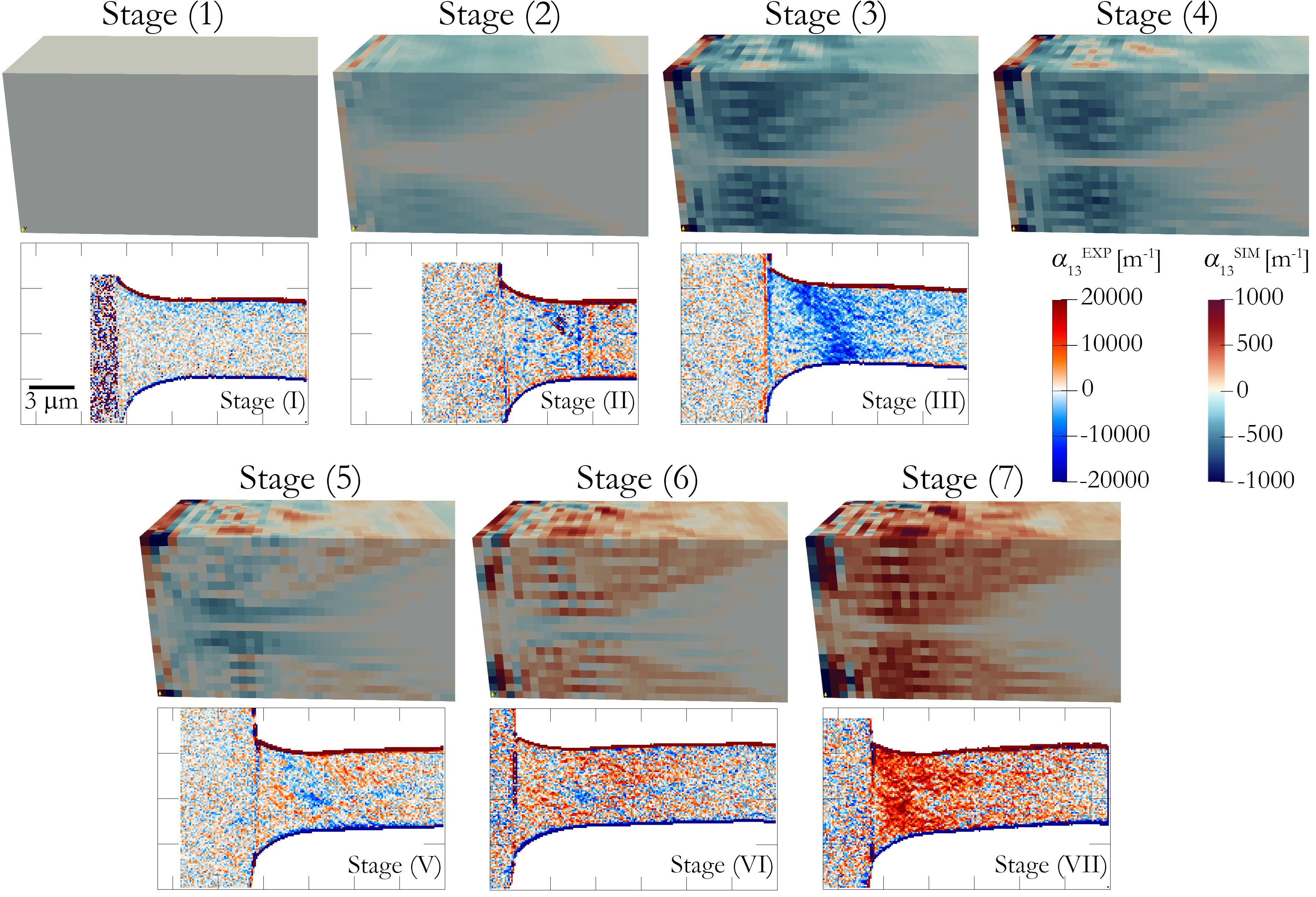}
    \caption{Evolution of $\alpha_{13}$ over the loading stages 1-7.\label{fig:alpha_evolution_sim}
    }
\end{figure*}

The evolution of the $\alpha_{13}$ component over the stages 1-7 is given in Fig.~\ref{fig:alpha_evolution_sim}. This shows the reversal of the signs, i.e., the change in polarisation during the evolution as required by the reversed strain gradients.

Finally, it shall be emphasised that the results show a significant Bauschinger effect during reverse loading based on the dislocation density evolution. The results show the formation, stabilisation and resolution of dislocation pileups as a result of the complex interplay between the different slip systems and their stress interaction.

\section{Discussion}
\label{sec:discussion}

In this paper a systematic analysis of collective dislocation phenomena has been performed for microcantilever bending. The experimental investigations are mostly based on the HR-EBSD technique that, as shown above, provides an exceptionally detailed access to the dislocation microstructure. It does not only deliver, among others, the spatial distribution of the GND density close to the sample's surface, but also three individual components of the Nye tensor that can be used to estimate the spatial distribution of BV types and their signs. We emphasise, that such a detailed analysis of the evolution of the dislocation microstructure was not possible before. Yet, this information, although very rich compared to other available methods, is far from complete. For instance, the microstructure at the surface (where the measurements are carried out) may differ from that of the sample interior and HR-EBSD cannot distinguish between the slip planes the dislocations are lying on. Therefore, the experiments were complemented by CDD simulations, that in principle are capable of modelling the evolution of the complete dislocation microstructure in the microsample, and provides access to all these missing information. As such, they help in understanding and evaluating the experimental results and the physics behind cantilever bending.

Our most important finding is related to the formation of the dislocation pile-ups and their irreversibility during bending. As explained in detail in the introduction, it has been argued that on the micron-scale dislocation pile-ups are often reversible, that is, they practically dissolve upon unloading. While this may be true for samples significantly smaller than studied here or for different orientations, here a completely different picture is found. During the first bending phase a structure similar to a set of simple planar pile-ups form, but when it comes to reversed loading the structure becomes highly complex with the GND density being distributed all along the microcantilever cross section and its base. The reason is, that the GNDs cannot fully escape the sample upon unloading, rather the majority of them gets stuck in the sample due to dislocation reactions, as seen in the experiments (bottom panel of Fig.~\ref{fig:evolution_exp}) and simulations (Fig.~\ref{fig:cycleloading}) as well. This is evident from the evolution of the available components of the tensor $\alpha$, too, where the change in sign (that is, polarisation) initiates from the top and bottom parts of the cantilever, and reaches its neutral axis only at larger reversed strains (see Fig.~\ref{fig:alpha_evolution_sim}) This means, that GNDs at the neutral axis are rather stable, e.g.,\ due to cross-slip or other short-range dislocation effects which produce junctions and keep them pinned. Nonetheless, as the reversed strain increases, the sign of the Nye tensor components ultimately change, and the structure gets polarised in the opposite direction. This process explains the strong strain hardening and the Bauschinger effect observed with both methods. The first one is the result of the accumulation of a large dislocation content due to reactions (a large portion of which are GNDs) and the other is due to the strong polarisation of the microstructure and related internal stresses, as proposed earlier \cite{mughrabi1988dislocation}.

It is important to emphasise the correspondence between experiments and CDD simulations. Of course a one-by-one comparison is not possible due to some differences between the two systems (such as the different geometry at the cantilever base, slight misorientation, imperfections of the sample and loading device, etc.) and due to the lower bending amplitudes applied in the simulations but the qualitative similarities between the results are evident. Not only the mechanical response and the GND density evolution, but also the evolution of the available Nye-tensor components show good accordance. In addition, both experiments and simulations predict that predominantly dislocations of Group I and II dominate the plastic response. Thus, CDD yields here a deeper insight into the dislocation mechanisms that play a role during the bending process.

The results demonstrate how different the plastic deformation is during the first and second half cycle of the microbending. During the first half cycle a fully polarised rather simple GND structure forms which upon reverse loading gets excessively complex sue to the processes described in the previous paragraph. This phenomenon may play a key role in the formation of dislocation patterns specific for fatigue (such as ladder structures in persistent slip bands).

The present study also pushes forward the applicability of HR-EBSD to investigate the evolution of the microstructure during plastic deformation in several aspects. (i) We validated the HR-EBSD measurements with TEM analysis and also showed their robustness by proving that the results were not sensitive to elastic unloading and a slight FIB polishing of the surface. (ii) The qualitative similarities with the CDD underline the physical feasibility of the HR-EBSD results. (iii) Although  HR-EBSD measurements can only be performed on the cantilever surface, CDD simulations can supplement our conclusions about the processes within the cantilever.

(iv) The applied BV analysis is consistent with the CDD simulations which has broadened the capabilities of HR-EBSD measurements and calls for further applications on materials with more complex microstructure.

In conclusion, in the considered size regime, we provided an in-depth \emph{in situ} analysis of how irreversible pile-up formation leads to the development of a complex polarised dislocation microstructure during cyclic loading. This approach opens new perspectives on understanding other microscopic mechanisms during fatigue and, thus, will assist  improving corresponding engineering design concepts.

\section*{Data availability}

Experimental data related to the HR-EBSD measurements and scripts for preparing figures of the manuscript are available at \url{https://doi.org/10.5281/zenodo.7970022}.

\section*{Declaration of Competing Interest}

The authors declare that they have no known competing financial interests or personal relationships that could have appeared to influence the work reported in this paper.

\section*{Acknowledgements}

Financial support from the National Research, Development and Innovation Fund of Hungary is acknowledged under the young researchers’ excellence programme NKFIH-FK-138975 (D.U., K.L., I.G.~and P.D.I), under the bilateral S\&T cooperation 2019-2.1.11-TÉT-2020-00173 (D.U.~and P.D.I.) and under the ÚNKP-21-4 New National Excellence Programme of the Ministry for Innovation and Technology (D.U.).
S.K.~was funded by the French National Research Agency (ANR) under the project No.~``ANR-22-CE08-0012-01" (INSTINCT). This paper was also supported by the János Bolyai Research Scholarship (Zs.F.) of the Hungarian Academy of Sciences. The authors would like to thank L.~Illés for the FIB preparation of the TEM lamella.
Furthermore, the financial support of this work in the context of the German Research Foundation (DFG) 
project SCHU 3074/3-1 is gratefully acknowledged. The simulation work
was performed on the computational resource HoreKa funded by the
Ministry of Science, Research and the Arts Baden-Württemberg and
DFG, Germany.

\section*{Appendix. Supplementary Material}

Supplementary Figures.

\bibliography{mybibfile}

\clearpage
\onecolumn
\section{Supplementary Figures}

\setcounter{figure}{0}
\renewcommand{\figurename}{Supplementary Figure}
\renewcommand{\thefigure}{S\arabic{figure}}

\begin{figure}[!h]
    \centering
    \includegraphics[width=0.48\textwidth]{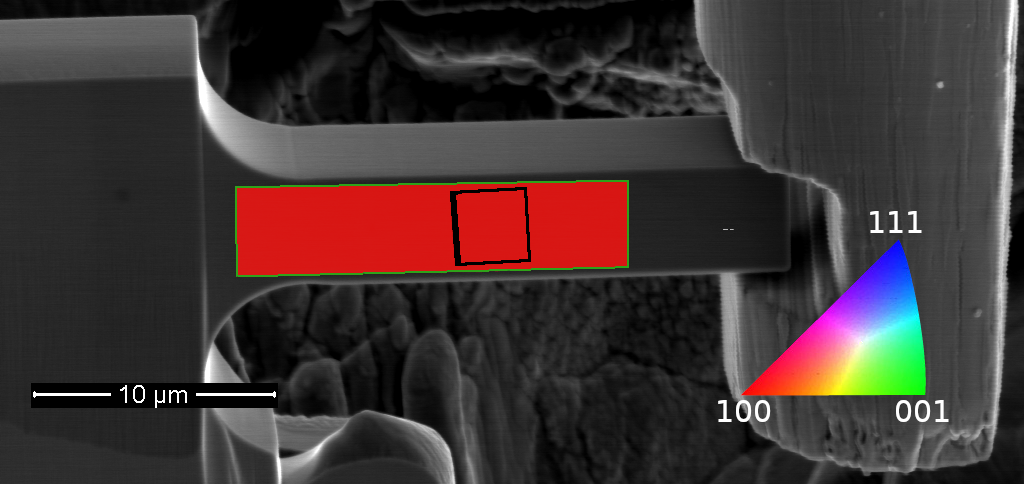}
    \caption{The initial EBSD map of the cantilever identifies the orientation. The depicted unit cell shows the geometry from the point of view perpendicular to the side surface of the cantilever.}
    \label{fig:C3-EBSD-orient}
\end{figure}

\begin{figure}[!h]
    \centering
    \includegraphics[width=0.48\textwidth]{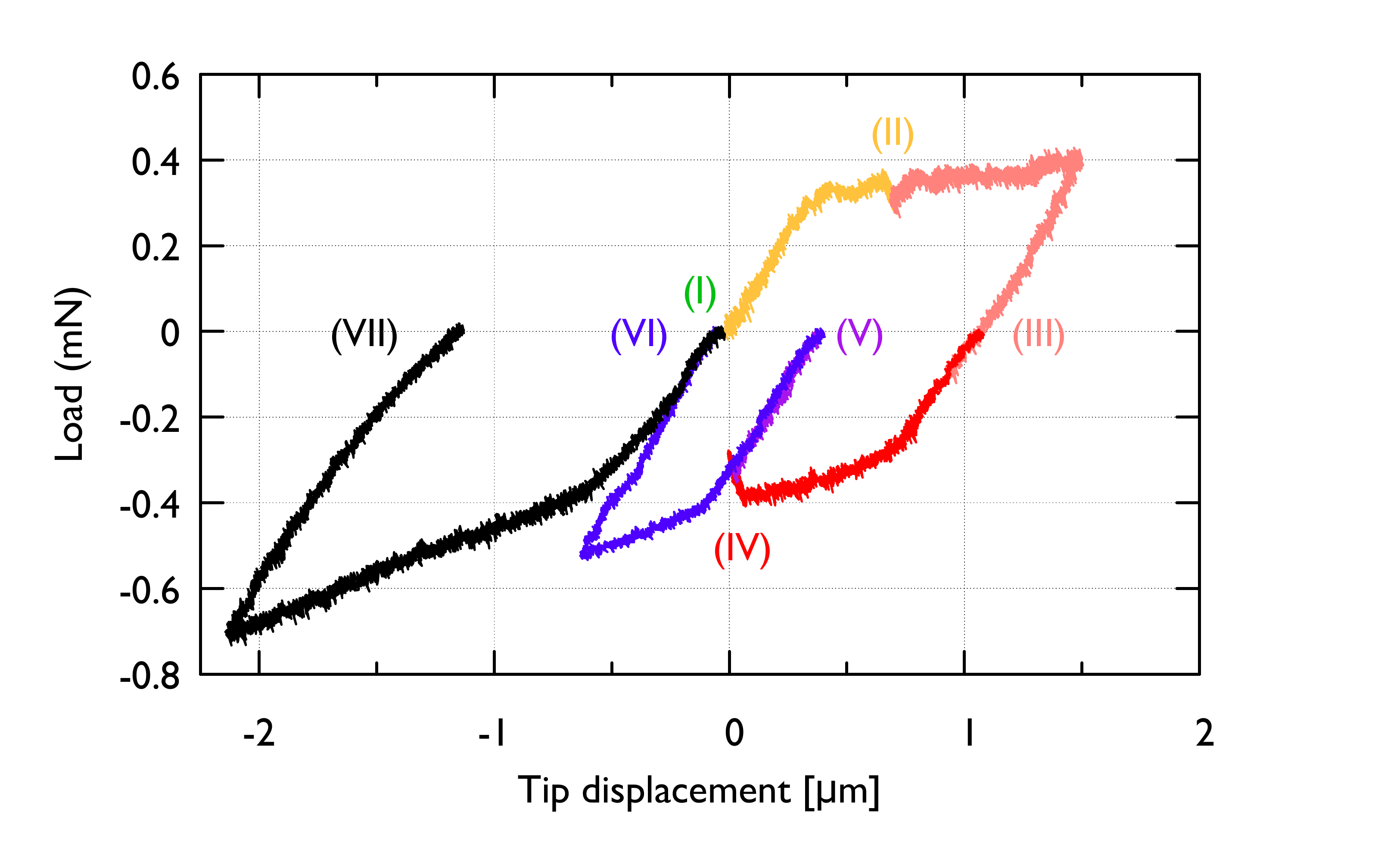}
    \caption{Load-displacement curve of the microcantilever subject to bi-directional bending.}
    \label{fig:C3-ld2}
\end{figure}

\begin{figure*}[!h]
    \centering
    \includegraphics[width=0.4\textwidth]{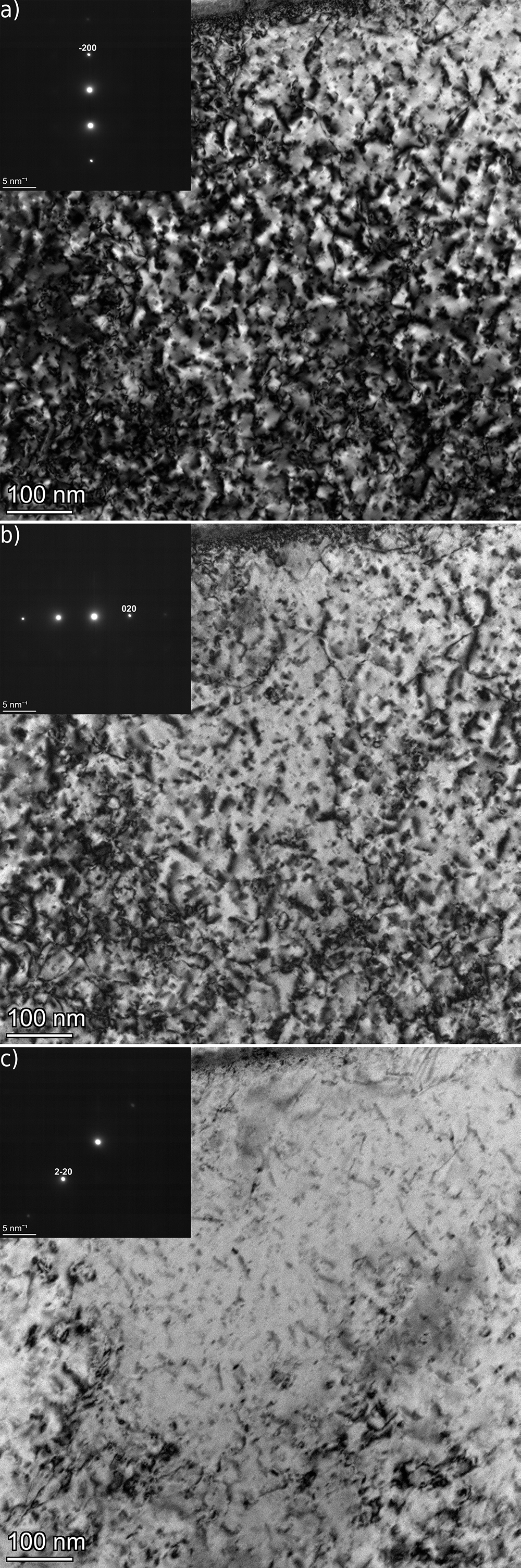}
    \caption{TEM results. a) BF in (200), b) (020), c) $(2\bar{2}0)$ two-beam conditions.}
    \label{fig:tem_supp}
\end{figure*}

\begin{figure*}[!h]
    \centering
    \includegraphics[width=0.8\textwidth]{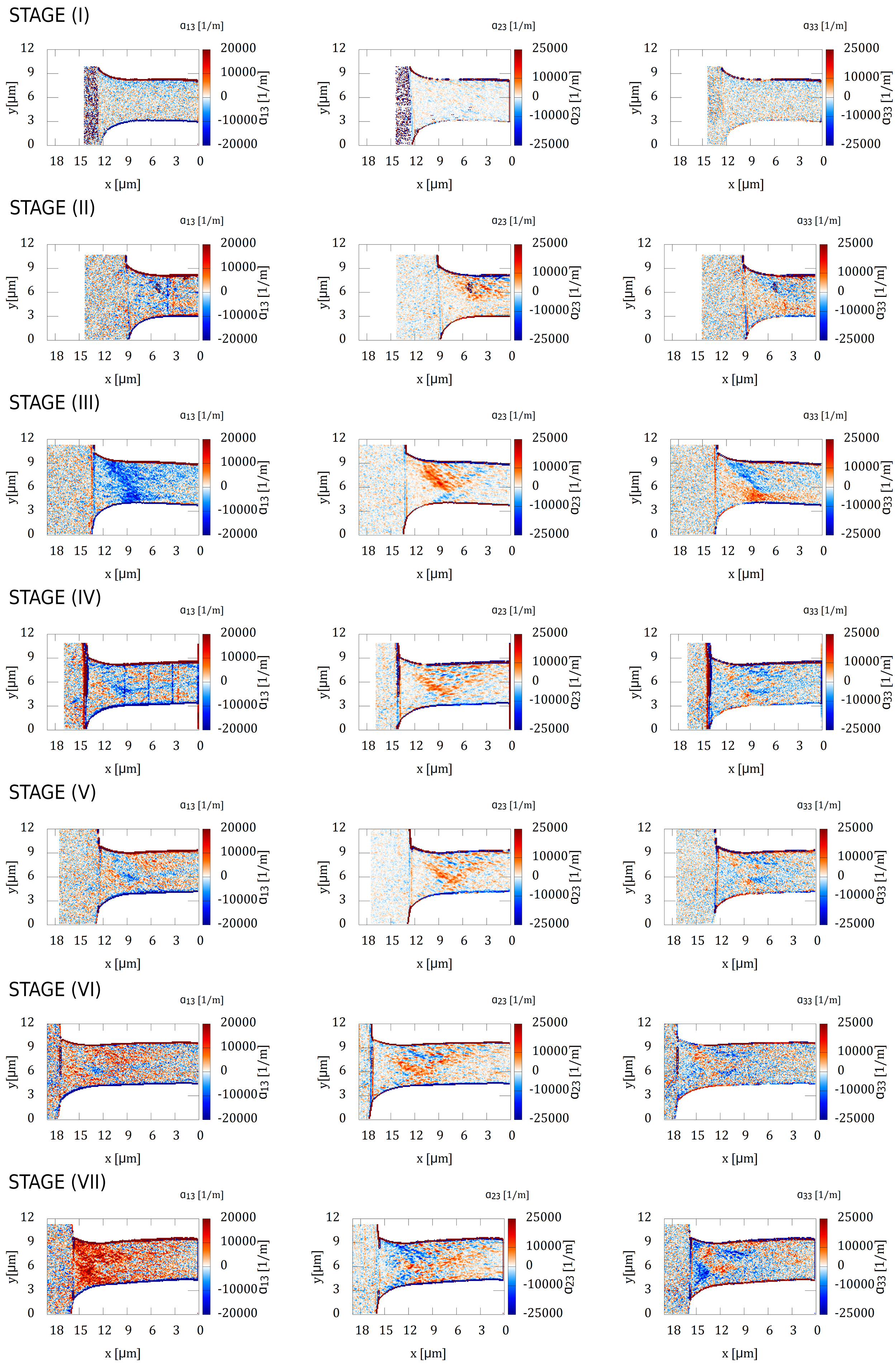}
    \caption{$\alpha_{i3}$ components.}
    \label{fig:alpha_evolution}
\end{figure*}

\begin{figure*}[!h]
    \centering
    \includegraphics[width=0.99\textwidth]{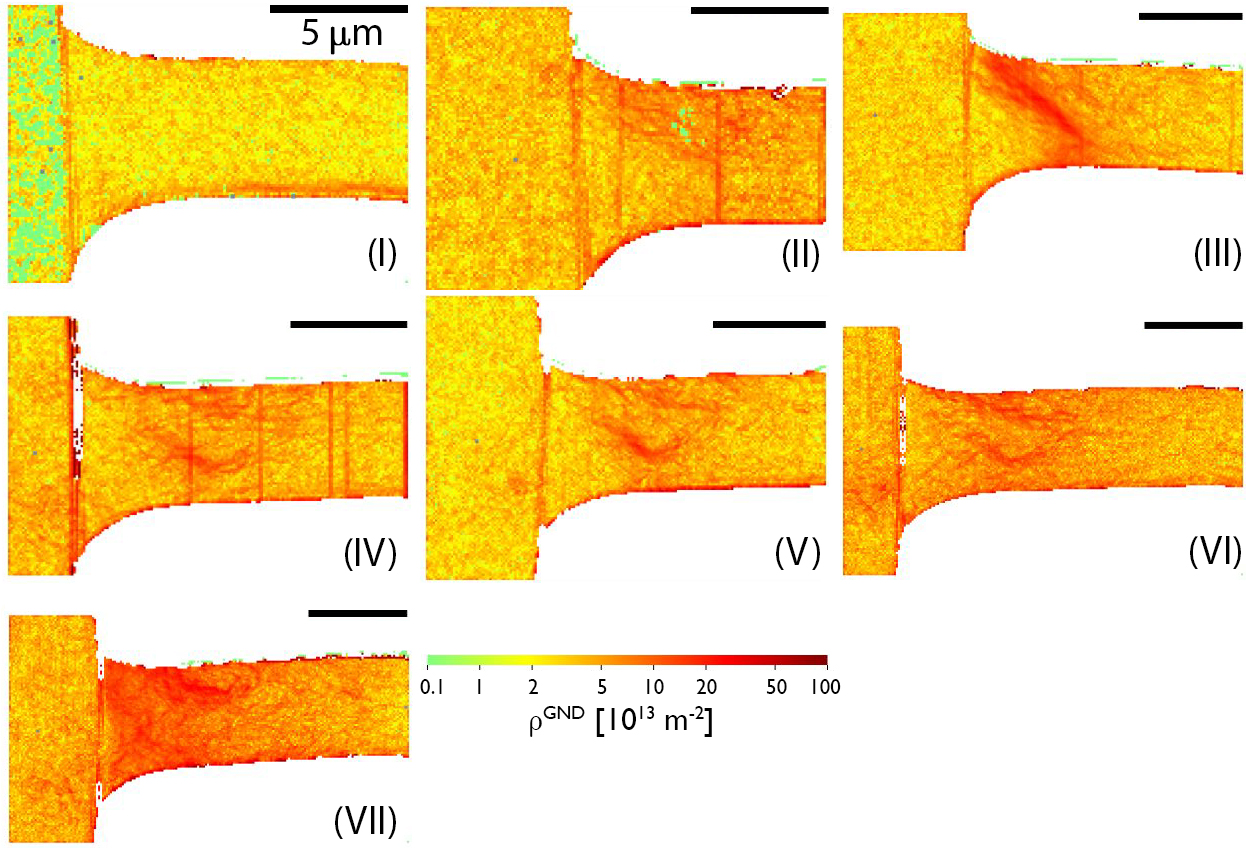}
    \caption{GND density maps. Scale bars correspond to 5 $\upmu$m.}
    \label{fig:GND_evolution}
\end{figure*}

\begin{figure}[!h]
    \centering
    \includegraphics[width=0.4\textwidth]{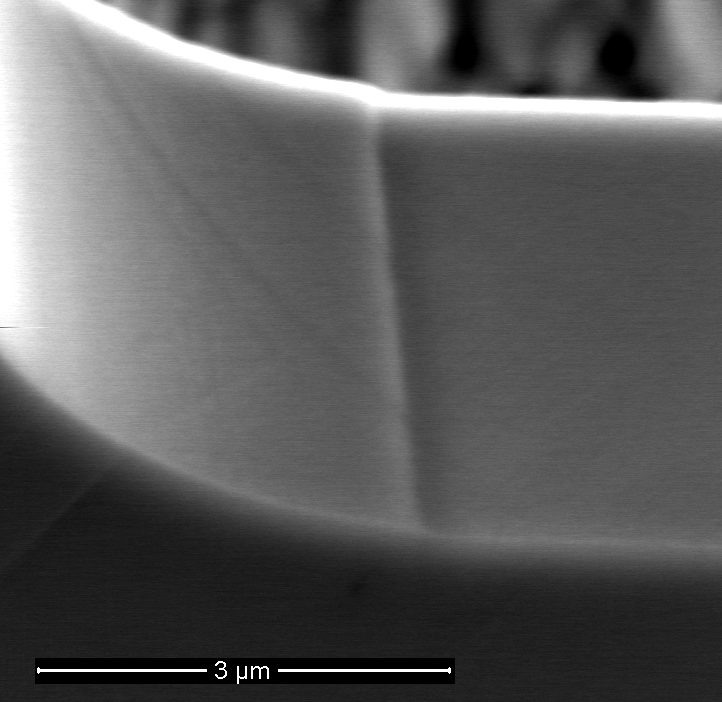}
    \caption{SEM image shows the formed shear bands on the cantilever's top side right after the first unloading.}
    \label{fig:SEM-pattern}
\end{figure}

\end{document}